\documentstyle[12pt,emlines]{article}
\newcommand{\dfrac}{\displaystyle \frac}
\newcommand{\nb}[1]{\mbox{\normalsize #1}}
\def\ao{{}\kern-.10em\hbox{``}}
\def\lint{\int\limits}
\begin{document}
\large
\bibliographystyle{plain}

\begin{titlepage}
\hfill \begin{tabular}{l} HEPHY-PUB 594/93\\ UWThPh-1993-59\\ December 1993
\end{tabular}\\[4cm]
\begin{center}
{\Large\bf EFFECTIVELY SEMI-RELATIVISTIC HAMILTONIANS OF NONRELATIVISTIC
FORM}\\
\vspace{1.5cm}
{\Large\bf Wolfgang LUCHA}\\[.5cm]
{\large Institut f\"ur Hochenergiephysik\\
\"Osterreichische Akademie der Wissenschaften\\
Nikolsdorfergasse 18, A-1050 Wien, Austria}\\[1cm]
{\Large\bf Franz F. SCH\"OBERL} {\Large and}
{\Large\bf Michael MOSER}\\[.5cm]
{\large Institut f\"ur Theoretische Physik\\
Universit\"at Wien\\
Boltzmanngasse 5, A-1090 Wien, Austria}\\[1.5cm]
{\bf Abstract}
\end{center}
\normalsize

We construct effective Hamiltonians which despite their apparently
nonrelativistic form incorporate relativistic effects by involving parameters
which depend on the relevant momentum. For some potentials the corresponding
energy eigenvalues may be determined analytically. Applied to two-particle
bound states, it turns out that in this way a nonrelativistic treatment may
indeed be able to simulate relativistic effects. Within the framework of
hadron spectroscopy, this lucky circumstance may be an explanation for the
sometimes extremely good predictions of nonrelativistic potential models even
in relativistic regions.
\end{titlepage}

\section{Introduction}

The fundamental disadvantage inherent to any (semi-) relativistically
consistent description of some quantum-theoretic system is obviously brought
about by the nonlocality of the \ao square-root" operator of the
relativistically correct kinetic energy, $\sqrt{\vec p\,{}^2 + m^2}$,
entering necessarily in the Hamiltonian $H$ which governs the dynamics of the
system under consideration. In contrast to the nonrelativistic limit,
obtained from the expansion of the square root up to the lowest $\vec
p\,{}^2$-dependent order, $\sqrt{\vec p\,{}^2 + m^2} = m + \vec
p\,{}^2/(2\,m) + \dots$, the presence of the relativistic kinetic-energy
operator prevents, in general, a thoroughly analytic discussion; one is
forced to rely on some numerical solution of the problem.

This inconvenience may be circumvented---at least in principle---by
approximating a given semi-relativistic Hamiltonian $H$ (incorporating, by
definition, relativistic kinematics) by the corresponding \ao effectively
semi-relativistic" Hamiltonian, formulated and investigated according to the
lines proposed in the present work. These effective Hamiltonians are
characterized by their rigorous maintenance of the easier to handle
nonrelativistic kinematics while resembling the relativistic formalisms to
the utmost possible extent by replacing their intrinsic parameters by
effective ones which depend in a well-defined manner on the square of the
momentum $\vec p$.

In order to be as concrete as possible, we choose to illustrate our route of
constructing and evaluating these effectively semi-relativistic Hamiltonians
for the particular case of bound states of two particles of spin zero. For
simplicity, let us assume that the two constituents of these bound states are
of equal mass $m$; the generalization to different masses is then
straightforward. In the framework of a semi-relativistic description all the
forces acting between these two particles may be derivable from some
coordinate-dependent interaction potential $V(\vec x)$. Consequently, the
semi-relativistic Hamiltonian describing this system in the
center-of-momentum frame of its constituents is given by
\begin{equation}
H = 2\sqrt{\vec p\,{}^2 + m^2} + V(\vec x) \quad .
\label{eq:hsemi-rel}
\end{equation}

The equation of motion resulting from this type of Hamiltonian is usually
called \ao spinless Salpeter equation." As it stands, it represents a
standard approximation to the Bethe--Salpeter formalism for bound states
within a relativistic quantum field theory. It is derived from the
Bethe--Salpeter equation \cite{salpeter51}
\begin{enumerate}
\item by eliminating---in accordance with the spirit of an instantaneous
interaction---any dependence on timelike variables, which leads to the
so-called \ao Salpeter equation" \cite{salpeter52}, and
\item by neglecting any reference to the spin degrees of freedom of the two
involved bound-state constituents and restricting to solutions corresponding
exclusively to positive energy.
\end{enumerate}

The outline of this paper is as follows. We introduce in Sect.
\ref{sec:effsemrelham} the effectively semi-relativistic Hamiltonians
corresponding to the really semi-relativistic Hamiltonians $H$ of Eq.
(\ref{eq:hsemi-rel}) in their most general form and derive in Sect.
\ref{sec:genstrat} for the special case of power-law potentials some sort of
\ao master equation" for that central quantity the knowledge of which enables
us to imitate the effects of relativistic kinematics within a formally
nonrelativistic framework, namely, the expectation value of the square of the
momentum $\vec p$. From the consideration of the most important prototypes of
interaction potentials in Sect. \ref{sec:effhamappl} we are led to conclude,
in Sect. \ref{sec:effhamconcl}, that our effective Hamiltonians represent
indeed a viable alternative to the original semi-relativistic Hamiltonians
(\ref{eq:hsemi-rel}).

\section{Effectively Semi-Relativistic Hamiltonians}\label{sec:effsemrelham}

The main idea of our way of constructing effectively semi-relativistic
Hamiltonians has already been sketched in Refs. \cite{lucha91,lucha92}. The
starting point of this construction is a trivial but nevertheless fundamental
inequality. This inequality relates the expectation values, taken with
respect to (at this stage) arbitrary Hilbert-space vectors $|\rangle$
normalized to unity, of both the first and second powers of an Hermitian (or,
to be more precise, self-adjoint) but otherwise arbitrary operator ${\cal O}
= {\cal O}^\dagger$; it reads
$$
|\langle{\cal O}\rangle| \le \sqrt{\langle{\cal O}^2\rangle} \quad .
\label{eq:ineqop}
$$
Application of the above inequality to the relativistic kinetic-energy
operator $\sqrt{\vec p\,{}^2 + m^2}$ yields
$$
\left\langle\sqrt{\vec p\,{}^2 + m^2}\right\rangle
\le \sqrt{\langle\vec p\,{}^2\rangle + m^2} \quad .
$$
By employing this inequality, we obtain for the expectation value $\langle
H\rangle$ of the semi-relativistic Hamiltonian $H$, Eq. (\ref{eq:hsemi-rel}),
\begin{eqnarray}
\langle H\rangle
&=& 2\left\langle\sqrt{\vec p\,{}^2 + m^2}\right\rangle + \langle V\rangle
\le 2\,\sqrt{\langle\vec p\,{}^2\rangle + m^2} + \langle V\rangle \nonumber\\
&=& 2\,\frac{\langle\vec p\,{}^2\rangle + m^2}
{\sqrt{\langle\vec p\,{}^2\rangle + m^2}} + \langle V\rangle
= \left\langle 2\,\frac{\vec p\,{}^2 + m^2}
{\sqrt{\langle\vec p\,{}^2\rangle + m^2}} + V\right\rangle \quad .
\label{eq:hamexpectval}
\end{eqnarray}

{}From now on we specify the Hilbert-space vectors in all expectation values to
be the eigenstates of our Hamiltonian $H$. In this case the expectation value
of $H$, $\langle H\rangle$, as appearing, e.~g., in (\ref{eq:hamexpectval}),
becomes the corresponding semi-relativistic energy eigenvalue $E$, i.~e., $E
\equiv \langle H\rangle$, and the inequality (\ref{eq:hamexpectval}) tells us
that this energy eigenvalue is bounded from above by
$$
E \le \left\langle 2\,\frac{\vec p\,{}^2 + m^2}
{\sqrt{\langle\vec p\,{}^2\rangle + m^2}} + V\right\rangle \quad .
$$
The operator within brackets on the right-hand side of this inequality is an
\ao effectively semi-relativistic" Hamiltonian $H_{\nb{eff}}$ which
possesses, quite formally, the structure of a nonrelativistic Hamiltonian,
\begin{equation}
H_{\nb{eff}}
\equiv 2\,\frac{\vec p\,{}^2 + m^2}{\sqrt{\langle\vec p\,{}^2\rangle + m^2}}
+ V
= 2\,\hat m + \frac{\vec p\,{}^2}{\hat m} + V_{\nb {eff}} \quad ,
\label{eq:heffective}
\end{equation}
but involves, however, the effective mass
\begin{equation}
\hat m = \frac{1}{2}{\sqrt{\langle\vec p\,{}^2\rangle + m^2}}
\label{eq:meffective}
\end{equation}
and the effective nonrelativistic potential
\begin{equation}
V_{\nb{eff}} = \frac{2\,m^2}{\sqrt{\langle\vec p\,{}^2\rangle + m^2}}
- \sqrt{\langle\vec p\,{}^2\rangle + m^2} + V
= 2\,\hat m - \frac{\langle\vec p\,{}^2\rangle}{\hat m} + V \quad .
\label{eq:poteffective}
\end{equation}
The effective mass $\hat m$ as given by Eq. (\ref{eq:meffective}) as well as
the constant, i.~e., coordinate-independent, term in the effective potential
$V_{\nb {eff}}$ of Eq. (\ref{eq:poteffective}), $2\,\hat m - \langle\vec
p\,{}^2\rangle/\hat m$, obviously depend on the expectation value of the
square of the momentum $\vec p$, $\langle\vec p\,{}^2\rangle$, and will
therefore differ for different energy eigenstates.

Motivated by our above considerations, we propose to approximate the true
energy eigenvalues $E$ of the semi-relativistic Hamiltonian $H$ of Eq.
(\ref{eq:hsemi-rel}) by the corresponding \ao effective" energy eigenvalues
$E_{\nb{eff}}$, defined as the expectation values of some effective
Hamiltonian $\tilde H_{\nb{eff}}$ taken with respect to the eigenstates
$|\rangle_{\nb{eff}}$ of its own,
$$
E_{\nb{eff}} = \langle\tilde H_{\nb{eff}}\rangle_{\nb{eff}} \quad ,
$$
where the effective Hamiltonian $\tilde H_{\nb{eff}}$, as far as its form is
concerned, is given by Eqs. (\ref{eq:heffective}) to (\ref{eq:poteffective})
but is implicitly understood to involve the expectation values of $\vec
p\,{}^2$ with respect to the effective eigenstates $|\rangle_{\nb{eff}}$
(that is, $\langle\vec p\,{}^2\rangle_{\nb{eff}}$ in place of $\langle\vec
p\,{}^2\rangle$):
$$
\tilde H_{\nb{eff}} = 4\,\tilde m
+ \frac{\vec p\,{}^2 - \langle\vec p\,{}^2\rangle_{\nb{eff}}}{\tilde m} + V
\quad ,
$$
with
$$
\tilde m = \frac{1}{2}{\sqrt{\langle\vec p\,{}^2\rangle_{\nb{eff}} + m^2}}
\quad .
$$
Accordingly, the effective energy eigenvalues $E_{\nb{eff}}$ are given by a
rather simple formal expression, viz., by
\begin{equation}
E_{\nb{eff}} = 4\,\tilde m + \langle V\rangle_{\nb{eff}} \quad .
\label{eq:effenergy}
\end{equation}

\section{General Strategy of Evaluation}\label{sec:genstrat}

We intend to elaborate our general prescription for the construction of
effectively semi-relativistic Hamiltonians $\tilde H_{\nb{eff}}$ in more
detail for the particular case of power-law potentials depending only on the
radial coordinate $r \equiv |\vec x|$, i.~e., for potentials of the form
$V(r) = a\,r^n$ with some constant $a$. The reason for this restriction is
twofold:
\begin{enumerate}
\item On the one hand, for power-law potentials the virial theorem
\cite{lucha89,lucha90mpla} in its nonrelativistic form \cite{lucha91,lucha92}
appropriate for the present case,
$$
\left\langle\dfrac{\vec p\,{}^2}{\tilde m}\right\rangle_{\nb{eff}}
= \dfrac{1}{2}\left\langle r\,\dfrac{dV(r)}{dr}\right\rangle_{\nb{eff}} \quad ,
\label{eq:virial}
$$
enables us to replace the expectation value of the potential in
(\ref{eq:effenergy}) by a well-defined function of the expectation value of
the squared momentum:
$$
a\,\langle r^n\rangle_{\nb{eff}}
= \dfrac{2}{n}\,\dfrac{\langle\vec p\,{}^2\rangle_{\nb{eff}}}{\tilde m} \quad .
$$
This implies for the effective energy eigenvalues
\begin{equation}
E_{\nb{eff}} = 4\,\tilde m + \dfrac{2}{n}\,
\dfrac{\langle\vec p\,{}^2\rangle_{\nb{eff}}}{\tilde m} \quad .
\label{eq:eeff}
\end{equation}
\item On the other hand, we may take advantage of the fact that for power-law
potentials it is possible to pass, without change of the fundamental
commutation relations between coordinate variables and their canonically
conjugated momenta, from the dimensional phase-space variables employed at
present to new, dimensionless phase-space variables and to rewrite the
Hamiltonian in form of a Hamiltonian which involves only these dimensionless
phase-space variables \cite{lucha91}. The eigenvalues $\epsilon$ of this
dimensionless Hamiltonian are, of course, also dimensionless \cite{lucha91}.
Applying this procedure, we find for the effective energy eigenvalues
$$
E_{\nb{eff}} - 4\,\tilde m
+ \dfrac{\langle \vec p\,{}^2\rangle_{\nb{eff}}}{\tilde m}
= \left\langle\dfrac{\vec p\,{}^2}{\tilde m} + a\,r^n\right\rangle_{\nb{eff}}
= \left(\dfrac{a^2}{{\tilde m}^n}\right)^{\frac{1}{2+n}} \epsilon \quad .
$$
\end{enumerate}
Combining both of the above expressions for $E_{\nb{eff}}$, we obtain a
relation which allows us to determine $\langle\vec p\,{}^2\rangle_{\nb{eff}}$
unambiguously in terms of the dimensionless energy eigenvalues $\epsilon$:
\begin{equation}
\langle\vec p\,{}^2\rangle_{\nb{eff}}^{2+n}
= \dfrac{1}{4}\left(\dfrac{n}{2+n}\right)^{2+n}a^2\,\epsilon^{2+n}\,
(\langle\vec p\,{}^2\rangle_{\nb{eff}} + m^2) \quad .
\label{eq:master}
\end{equation}
For a given power $n$ this equation may be solved for $\langle\vec
p\,{}^2\rangle_{\nb{eff}}$. Insertion of the resulting expression into Eq.
(\ref{eq:eeff}) then yields the corresponding eigenvalue $E_{\nb{eff}}$ of
the effectively semi-relativistic Hamiltonian $\tilde H_{\nb{eff}}$.

\section{Applications}\label{sec:effhamappl}

We would like to investigate the capabilities of the effective treatment
proposed in the previous sections by discussing some of its implications for
some familiar prototypes of interaction potentials, namely, for the
harmonic-oscillator, Coulomb, linear, and funnel potential. To this end we
compare for the lowest-lying energy eigenstates (which we will label
according to the usual spectroscopic notation) the energy eigenvalues
$E_{\nb{eff}}$ resulting from our effective description with the respective
energy eigenvalues $E_{\nb{NR}}$ obtained within the corresponding and by now
rather standard nonrelativistic approach \cite{lucha91,lucha92}. Although for
obvious reasons we do not intend at this place to perform a fit of some
experimentally observed particle spectrum, we employ, for the purpose of
comparison, a set of numerical values of the involved parameters which
represents the typical orders of magnitude of the various constants found
within a phenomenological description of hadrons as bound states of quarks by
(nonrelativistic) potential models \cite{lucha91,lucha92}. In particular, our
choice for the mass $m$ of the bound-state constituents is $m = 1.8$ GeV,
which corresponds to the typical mass of the constituent c quark.

Harmonic-oscillator and Coulomb potential may be investigated on purely
algebraic grounds. More sophisticated potentials, however, have to be handled
with the help of numerical methods \cite{falk85}. Occasionally, it will prove
to be favourable to inspect in particular the ultrarelativistic limit of the
developed formalism, defined by vanishing mass $m$ of the bound-state
constituents, i.~e., by $m = 0$.

An important feature of the experimentally measured mass spectra of
hadrons---which may serve to provide a decisive criterion regarding the
usefulness of our effective treatment for a meaningful description of
hadrons---is the empirically well-established linearity of the Regge
trajectories: both mesons and baryons may be grouped to form sets of
particles which populate (approximately) linear Regge trajectories; the
different members of these sets are related by the fact that, apart from a
constant shift, the squares of their masses, i.~e., of the energy eigenvalues
of the corresponding bound states of quarks in their center-of-momentum
frame, are proportional to the relative orbital angular momentum $\ell$ of
the bound-state constituents or, equivalently, the spin of the composite
particles, with almost one and the same constant of proportionality, the
so-called Regge slope $\beta \simeq 1.2 \mbox{ GeV}^2$, for all Regge
trajectories \cite{pdg92}. We indicate these relationships by $E^2(\ell) =
\beta\,\ell + \mbox{const}$. In general, the theoretical dependence of the
energy eigenvalues $E$ on the angular momentum $\ell$ will turn out to be
described by some rather complicated function of $\ell$. For this reason we
only take a quick glance on the asymptotic behaviour of the predicted energy
eigenvalues $E(\ell)$ for large values of the angular momentum $\ell$,
symbolically denoted by the limit $\ell \to \infty$. There we may expect to
observe a simple power-law rise of the calculated squares of energy
eigenvalues $E^2(\ell)$ for increasing values of $\ell$.

\subsection{Harmonic oscillator}

For the harmonic-oscillator potential $V(r) = a\,r^2$, that is, for $n = 2$,
Eq. (\ref{eq:master}) reduces to a quartic equation for the expectation value
$\langle\vec p\,{}^2\rangle_{\nb{eff}}$. Inserting the well-known expression
\cite{lucha91} for the dimensionless energy eigenvalues $\epsilon$ of the
three-dimensional harmonic oscillator, $\epsilon = 2\,N$, where $N$ is given
in terms of the radial and orbital angular-momentum quantum numbers $n_r$ and
$\ell$, respectively, by $N = 2\,n_r + \ell + \frac{3}{2}$, and introducing
the shorthand notation $k \equiv \sqrt{\frac{a}{2}}\,N$ and
$$
x \equiv \left(\frac{k^8}{2}\right)^\frac{1}{3}\left[
\left(1 + \sqrt{\dfrac{256}{27}\,\dfrac{m^6}{k^4} + 1}\right)^\frac{1}{3} +
\left(1 - \sqrt{\dfrac{256}{27}\,\dfrac{m^6}{k^4} + 1}\right)^\frac{1}{3}
\right] \quad ,
$$
the analytic solution of this quartic equation for $\langle\vec
p\,{}^2\rangle_{\nb{eff}}$ reads
$$
\langle\vec p\,{}^2\rangle_{\nb{eff}}
= \dfrac{\sqrt{x}}{2} + \sqrt{\dfrac{k^4}{2\sqrt{x}} - \dfrac{x}{4}} \quad .
$$
According to our above prescription, the effective energy eigenvalue is then
given by inserting this result into Eq. (\ref{eq:eeff}). In the
ultrarelativistic limit this effective energy eigenvalue takes a particularly
simple form: from $m = 0$ one finds $\langle\vec p\,{}^2\rangle_{\nb{eff}} =
k^\frac{4}{3}$ and $E_{\nb{eff}} = 4\sqrt{\langle\vec
p\,{}^2\rangle_{\nb{eff}}} = 2\,(4\,a)^\frac{1}{3}N^\frac{2}{3}$.

Table \ref{tab:harmosci} compares the nonrelativistic and effectively
semi-relativistic approaches for the harmonic oscillator. Our choice of
parameter values implies for the ground state $E_{\nb{NR}} < E_{\nb{eff}}$
whereas for the excited states $E_{\nb{NR}} > E_{\nb{eff}}$ holds.
{\normalsize
\begin{table}[hbt]
\begin{center}
\caption{Energy eigenvalues (in GeV) of nonrelativistic and effectively
semi-relativistic Hamiltonian for the harmonic-oscillator potential $V(r) =
a\,r^2$, with $a = 0.5 \mbox{ GeV}^3$.}\label{tab:harmosci}
\vspace{0.5cm}
\begin{tabular}{|c|c|c|}
\hline
&&\\[-1ex]
\multicolumn{1}{|c|}{State}&
\multicolumn{1}{c|}{\begin{tabular}{c}
Nonrelativistic\\ Hamiltonian\end{tabular}}&
\multicolumn{1}{c|}{\begin{tabular}{c}
Effective\\ Hamiltonian\end{tabular}}\\
&&\\[-1.5ex]
\hline
&&\\[-1.5ex]
$\quad\mbox{1S}\quad$&$\quad$5.181$\quad$&$\quad$5.198$\quad$\\
$\quad\mbox{1P}\quad$&$\quad$6.235$\quad$&$\quad$6.188$\quad$\\
$\quad\mbox{2S/1D}\quad$&$\quad$7.289$\quad$&$\quad$7.128$\quad$\\[1.5ex]
\hline
\end{tabular}
\end{center}
\end{table}}

Furthermore, it is no problem to determine immediately the large-$\ell$
behaviour of the theoretical energy eigenvalues. In the ultrarelativistic
case, because of $N \propto \ell$ for large $\ell$, the effective energy
eigenvalues $E_{\nb{eff}}$ behave, according to their above-mentioned
explicit general form, like $E_{\nb{eff}}^2(\ell) \propto \ell^\frac{4}{3}$.
In contrast to that, the large-$\ell$ asymptotic behaviour of the
corresponding nonrelativistic energy eigenvalues $E_{\nb{NR}}$ is given by
$E_{\nb{NR}} = 2\sqrt{\frac{a}{m}}\,\ell + \mbox{const.}$
\cite{lucha91,lucha92}, which implies $E_{\nb{NR}}^2(\ell) \propto \ell^2$.
We conclude that for the harmonic-oscillator potential (at least the
ultrarelativistic limit of) the effective treatment comes closer to the
observed linearity of the Regge trajectories than the nonrelativistic
approach.

\subsection{Coulomb potential}

For the Coulomb potential $V(r) = - \kappa/r$, that is, for $n = - 1$, Eq.
(\ref{eq:master}) reduces to a linear equation for the expectation value
$\langle\vec p\,{}^2\rangle_{\nb{eff}}$. Inserting the well-known expression
\cite{lucha91} for the dimensionless energy eigenvalues $\epsilon$ of the
Coulomb problem, $\epsilon = - (2\,N)^{- 2}$, where $N$ is given in terms of
the radial and orbital angular-momentum quantum numbers $n_r$ and $\ell$,
respectively, by $N = n_r + \ell + 1$, we obtain from this linear equation
for $\langle\vec p\,{}^2\rangle_{\nb{eff}}$
$$
\langle\vec p\,{}^2\rangle_{\nb{eff}}
= \dfrac{\kappa^2\,m^2}{16\,N^2 - \kappa^2} \quad ,
$$
and, after inserting this expression into Eq. (\ref{eq:eeff}), for the
effective energy eigenvalue
$$
E_{\nb{eff}}
= \dfrac{m}{N}\,\dfrac{8\,N^2 - \kappa^2}{\sqrt{16\,N^2 - \kappa^2}} \quad .
$$

For the case of vanishing orbital angular momentum, i.~e., for $\ell = 0$, an
analytic expression for the energy eigenvalues $E_{\nb{SR}}$ of the genuine
semi-relativistic Hamiltonian (\ref{eq:hsemi-rel}) may be derived, which
reads \cite{durand83}
$$
E_{\nb{SR}} = \dfrac{2\,m}{\sqrt{1 + \dfrac{\kappa^2}{4\,n^2}}} \quad , \qquad
n = 1,2,\dots \quad .
$$

Within both of the above approaches all energy eigenvalues vanish in the
ultrarelativistic limit $m = 0$. For the Coulomb problem, because of the lack
of any sort of dimensional parameter inherent to the theory in the case $m =
0$, this kind of degeneracy must take place already for dimensional reasons.
It may be understood completely by application of the general, that is,
relativistic, virial theorem \cite{lucha89,lucha90mpla} derived by two of the
present authors.

Table \ref{tab:coulomb} compares the nonrelativistic and effectively
semi-relativistic approaches for the Coulomb potential. For the numerical
values of our parameters employed at present all energy levels calculated
within the effective treatment surmount their counterparts in the
nonrelativistic description: $E_{\nb{eff}} > E_{\nb{NR}}$.
{\normalsize
\begin{table}[hbt]
\begin{center}
\caption{Energy eigenvalues (in GeV) of nonrelativistic and effectively
semi-relativistic Hamiltonian for the Coulomb potential $V(r) = - \kappa/r$,
with $\kappa = 0.456$.}\label{tab:coulomb}
\vspace{0.5cm}
\begin{tabular}{|c|c|c|}
\hline
&&\\[-1ex]
\multicolumn{1}{|c|}{State}&
\multicolumn{1}{c|}{\begin{tabular}{c}
Nonrelativistic\\ Hamiltonian\end{tabular}}&
\multicolumn{1}{c|}{\begin{tabular}{c}
Effective\\ Hamiltonian\end{tabular}}\\
&&\\[-1.5ex]
\hline
&&\\[-1.5ex]
$\quad\mbox{1S}\quad$&$\quad$3.5064$\quad$&$\quad$3.5294$\quad$\\
$\quad\mbox{2S/1P}\quad$&$\quad$3.5766$\quad$&$\quad$3.5824$\quad$\\
$\quad\mbox{3S/2P/1D}\quad$&$\quad$3.5896$\quad$&$\quad$3.5922$\quad$\\[1.5ex]
\hline
\end{tabular}
\end{center}
\end{table}}

Picking up the question of the large-$\ell$ behaviour of the theoretical
energy eigenvalues again, we find from the reported explicit expression that
in the limit $\ell \to \infty$ the effective energy eigenvalues
$E_{\nb{eff}}$ will not depend on the orbital angular momentum $\ell$ at all:
$E_{\nb{eff}}^2(\ell) \propto \ell^0$. In the nonrelativistic case, on the
other hand, the energy eigenvalues $E_{\nb{NR}}$ behave asymptotically like
$E_{\nb{NR}} = - \frac{m\,\kappa^2}{4\,\ell^2} + \mbox{const.}$
\cite{lucha91,lucha92}. Because of the negative sign in front of the
$\ell$-dependent term this yields a rise of the form $E_{\nb{NR}}^2(\ell)
\propto \ell^{-4}$ for increasing values of $\ell$.

\subsection{Variational method}\label{subsec:varmeth}

In general, it will not be possible to find some analytic expressions for the
effective energy eigenvalues $E_{\nb{eff}}$. However, in order to obtain an
approximation to the spectrum of energy eigenvalues to be expected or to get,
at least, some idea of it one may adopt the variational method described in
the following.

This standard variational method proceeds along the steps of the following,
extremely simple recipe \cite{lucha92,flamm82}:
\begin{enumerate}
\item Choose a suitable set of trial states $|\lambda\rangle$. The members of
this set are distinguished by some sort of variational parameter $\lambda$.
\item Compute the set of expectation values of the Hamiltonian under
consideration, $H$, with respect to these trial states $|\lambda\rangle$ in
order to obtain $E(\lambda) \equiv \langle\lambda|H|\lambda\rangle$.
\item Determine that value of the variational parameter $\lambda$---say,
$\lambda_{\nb{min}}$---which minimizes the resulting, $\lambda$-dependent
expression $E(\lambda)$.
\item Compute $E(\lambda)$ at the point of the minimum $\lambda_{\nb{min}}$
to find in this way the minimal expectation value $E(\lambda_{\nb{min}})$ of
the Hamiltonian $H$ in the Hilbert-space subsector of the chosen trial states
$|\lambda\rangle$.
\end{enumerate}
This minimum $E(\lambda_{\nb{min}})$ provides, of course, only an upper bound
to the proper energy eigenvalue of the Hamiltonian $H$.\footnote{\normalsize\
The accuracy of this method is discussed in Ref. \cite{schoeberl82}.}

Application of this straightforward variational procedure to one of our
effectively semi-relativistic Hamiltonians $\tilde H_{\nb{eff}}$ leads to
$E_{\nb{eff}}(\lambda_{\nb{min}})$, which, according to its derivation,
represents at least an upper bound to the corresponding effective energy
eigenvalue $E_{\nb{eff}}$.

Note that, as far as the above variational procedure is concerned, the
expectation value $\langle\vec p\,{}^2\rangle_{\nb{eff}}$ entering in the
effective Hamiltonian has to be regarded as a constant. Consequently, it has
not to be taken into account in the course of minimization of the energy
expression $E(\lambda)$ by varying the characteristic parameter $\lambda$.
Rather, in the framework of this variational technique, it has to be equated
to the expectation value of $\vec p\,{}^2$ taken with respect to precisely
that trial state $|\lambda_{\nb{min}}\rangle$ which is characterized by just
the minimizing value $\lambda_{\nb{min}}$ of the variational parameter
$\lambda$, that is, to $\langle\lambda_{\nb{min}}|\vec
p\,{}^2|\lambda_{\nb{min}}\rangle$.

For the present investigation we adopt the simplest conceivable set of trial
states $|\lambda\rangle$, namely, the ones the coordinate-space
representation $\psi(\vec x)$ of which is given, for vanishing radial quantum
number $n_r$, by the Gaussian trial functions (w.~l.~o.~g., $\lambda > 0$)
$$
\psi_{\ell m}(r,\theta,\phi)
= \sqrt{\dfrac{2\,\lambda^{2\,\ell + 3}}
{\Gamma\left(\ell + \mbox{$\frac{3}{2}$}\right)}}\,r^\ell\,
\exp\left(- \dfrac{\lambda^2\,r^2}{2}\right)
{\cal Y}_{\ell m}(\theta,\phi) \quad ,
$$
where ${\cal Y}_{\ell m}$ denote the spherical harmonics for angular momentum
$\ell$ and projection $m$, and the normalization factor of these trial
functions makes use of the so-called gamma function \cite{abramow}
$$
\Gamma(z) \equiv \lint_0^\infty dt\,t^{z - 1}\,\exp(- t) \quad .
$$
For this particular set of trial functions we obtain for the expectation
values of the square $\vec p\,{}^2$ of the momentum $\vec p$ and of the
$n$-th power $r^n$ of the radial coordinate $r$, respectively, with respect
to the trial states $|\lambda\rangle$
$$
\langle\lambda|\vec p\,{}^2|\lambda\rangle
= \left(\ell + \mbox{$\frac{3}{2}$}\right)\lambda^2
$$
and
$$
\langle\lambda|r^n|\lambda\rangle
= \dfrac{\Gamma\left(\ell + \mbox{$\frac{3 + n}{2}$}\right)}
{\Gamma\left(\ell + \mbox{$\frac{3}{2}$}\right)}\,
\dfrac{1}{\lambda^n} \quad .
$$

\subsection{Linear potential}

For the linear potential $V(r) = a\,r$, that is, for $n = 1$, Eq.
(\ref{eq:master}) reduces to a cubic equation for the expectation value
$\langle\vec p\,{}^2\rangle_{\nb{eff}}$ which, of course, may be solved
analytically. Unfortunately, for the linear potential the dimensionless
energy eigenvalues $\epsilon$ are only known \cite{lucha91} for the case of
vanishing orbital angular momentum $\ell$, i.~e., only for $\ell = 0$. In
this case they are given by the negative zeros of the Airy function
\cite{abramow}. In any case, that is, for arbitrary values of $\ell$, the
effective energy eigenvalues $E_{\nb{eff}}$ may be found by employing some
numerical procedure.

However, before performing a numerical computation of the energy eigenvalues
$E_{\nb{eff}}$ of the effective Hamiltonian $\tilde H_{\nb{eff}}$ with linear
potential, we apply the simple variational technique introduced in the
preceding subsection. For this Hamiltonian the value of the variational
parameter $\lambda$ which minimizes the expectation value
$\langle\lambda|\tilde H_{\nb{eff}}|\lambda\rangle$, that is,
$\lambda_{\nb{min}}$, is implicitly given by
$$
\lambda_{\nb{min}}^3 = \dfrac{a}{2}\,\dfrac{\Gamma(\ell + 2)}
{\Gamma\left(\ell + \mbox{$\frac{5}{2}$}\right)}\,\tilde m \quad .
$$
Recalling the definition of $\tilde m$ as given in Sect.
\ref{sec:effsemrelham}, we obtain from this expression a cubic equation for
$\langle\lambda_{\nb{min}}|\vec p\,{}^2|\lambda_{\nb{min}}\rangle$,
$$
\langle\lambda_{\nb{min}}|\vec p\,{}^2|\lambda_{\nb{min}}\rangle^3
= \dfrac{a^2}{16}\left(\ell + \mbox{$\frac{3}{2}$}\right)
\left(\dfrac{\Gamma(\ell + 2)}
{\Gamma\left(\ell + \mbox{$\frac{3}{2}$}\right)}\right)^2\left(
\langle\lambda_{\nb{min}}|\vec p\,{}^2|\lambda_{\nb{min}}\rangle + m^2\right)
\quad ,
$$
the analytic solution of which may be written down quickly. Insertion of this
result into Eq. (\ref{eq:eeff}) yields $E_{\nb{eff}}(\lambda_{\nb{min}})$ for
the linear potential. In the ultrarelativistic limit $m=0$ we find in this
way the (variational) effective energy eigenvalues
$$
E_{\nb{eff}}(\lambda_{\nb{min}})
= 3\left(\ell + \mbox{$\frac{3}{2}$}\right)^\frac{1}{4}
\sqrt{\dfrac{\Gamma(\ell + 2)}
{\Gamma\left(\ell + \mbox{$\frac{3}{2}$}\right)}\,a} \quad .
$$

Table \ref{tab:linear} compares the nonrelativistic and effectively
semi-relativistic approaches for the linear potential. According to our
announcement, for the effective treatment the energy eigenvalues have been
computed in an iterative way with the help of the numerical procedure
developed in Ref. \cite{falk85}. Again our choice of parameter values places
all effective energy levels above their nonrelativistic counterparts:
$E_{\nb{eff}} > E_{\nb{NR}}$. The variational calculation outlined above
yields, just for comparison, for the 1S and 1P state, respectively, the
(variational) effective energy eigenvalues $E_{\nb{eff}}(\lambda_{\nb{min}})
= 4.3107\mbox{ GeV}$ and $E_{\nb{eff}}(\lambda_{\nb{min}}) = 4.6168\mbox{
GeV}$, both of which are very close to the numerically obtained results and,
of course, upper bounds to them.
{\normalsize
\begin{table}[hbt]
\begin{center}
\caption{Energy eigenvalues (in GeV) of nonrelativistic and effectively
semi-relativistic Hamiltonian for the linear potential $V(r) = a\,r$, with $a
= 0.211\mbox{ GeV}^2$.}\label{tab:linear}
\vspace{0.5cm}
\begin{tabular}{|c|c|c|}
\hline
&&\\[-1ex]
\multicolumn{1}{|c|}{State}&
\multicolumn{1}{c|}{\begin{tabular}{c}
Nonrelativistic\\ Hamiltonian\end{tabular}}&
\multicolumn{1}{c|}{\begin{tabular}{c}
Effective\\ Hamiltonian\end{tabular}}\\
&&\\[-1.5ex]
\hline
&&\\[-1.5ex]
$\quad\mbox{1S}\quad$&$\quad$4.2812$\quad$&$\quad$4.3087$\quad$\\
$\quad\mbox{1P}\quad$&$\quad$4.5793$\quad$&$\quad$4.6149$\quad$\\
$\quad\mbox{2S}\quad$&$\quad$4.7911$\quad$&$\quad$4.8309$\quad$\\[1.5ex]
\hline
\end{tabular}
\end{center}
\end{table}}

Inspecting once again the large-$\ell$ behaviour of the predicted energy
eigenvalues, we may read off from the above explicit expression for the
ultrarelativistic (variational) effective energy eigenvalues
$E_{\nb{eff}}(\lambda_{\nb{min}})$, with the help of a useful relation
describing the asymptotic behaviour of the ratio of gamma functions
\cite{abramow}, viz.,
$$
\lim_{\ell \to \infty}\dfrac{\Gamma(\ell + z)}{\Gamma(\ell + u)}
= \ell^{z - u} \quad ,
$$
the very pleasing result $E_{\nb{eff}}^2(\lambda_{\nb{min}}) = 9\,a\,\ell$.
Accordingly, the effectively semi-relativistic energy eigenvalues of the
linear potential are perfectly able to reproduce the observed linearity of
the Regge trajectories with, however, a slope which is slightly larger than
the one obtained within different analyses based on the true
semi-relativistic Hamiltonian (\ref{eq:hsemi-rel}), all of which end up with
one and the same finding: $E_{\nb{SR}}^2 = 8\,a\,\ell$
\cite{kang75,lucha91regge}. Moreover, from the point of view of a correct
description of the linear Regge trajectories, both of the semi-relativistic
treatments are clearly superior to the corresponding nonrelativistic
approach, which gives for the energy eigenvalues of the linear potential
\cite{lucha91,lucha92}
$$
E_{\nb{NR}} =
3\left(\frac{a^2}{4\,m}\right)^\frac{1}{3}\ell^\frac{2}{3} + \mbox{const.}
$$
and therefore $E_{\nb{NR}}^2(\ell) \propto \ell^\frac{4}{3}$.

\subsection{Funnel potential}

Unfortunately, the potentials considered up to now are merely of more or less
academic interest. Finally, however, we would like to discuss a potential
which has been among the first ones to be proposed \cite{eichten75} for the
description of hadrons as bound states of constituent quarks, viz., the
funnel (or Cornell or Coulomb-plus-linear) potential.

This funnel potential comprehends the two basic ingredients of any realistic,
that is, phenomenologically acceptable, inter-quark potential, namely,
\begin{itemize}
\item at short inter-quark distances some Coulomb-like singularity of
perturbative origin, which arises from one-gluon exchange, and
\item at large inter-quark distances an approximately linear rise of
non-perturbative origin, which is responsible for colour confinement.
\end{itemize}
The funnel potential incorporates these two features in the simplest
conceivable manner:
$$
V(r) = - \dfrac{\kappa}{r} + a\,r \quad .
$$
In this form it still represents the prototype of almost all forthcoming
potential models designed to describe all the (binding) forces acting between
quarks.\footnote{\normalsize\ For a brief survey see, for instance, Ref.
\cite{lucha91}.}

This funnel potential is, beyond doubt, not of the power-law type.
Consequently, it cannot be subjected to the general effective formalism
developed so far but deserves a special treatment, which might consist of
some purely numerical approach.

However, as before, we first want to obtain some insight by applying the
variational procedure described in Subsect. \ref{subsec:varmeth}. The value
$\lambda_{\nb{min}}$ of the variational parameter $\lambda$ which minimizes
for the case of the above funnel potential the expectation value of the
effective Hamiltonian $\tilde H_{\nb{eff}}$ with respect to our Gaussian
trial states is (because of the presence of $\tilde m$ only implicitly)
determined by the relation
$$
\lambda_{\nb{min}}^3 = \dfrac{\tilde m}{2}\,
\dfrac{\Gamma(\ell + 1)}{\Gamma\left(\ell + \mbox{$\frac{5}{2}$}\right)}
\left[\kappa\,\lambda_{\nb{min}}^2 + a\,(\ell + 1)\right] \quad .
$$
In the ultrarelativistic limit $m = 0$ this relation fixes
$\lambda_{\nb{min}}$ to
$$
\lambda_{\nb{min}}
= \sqrt{\dfrac{a\,\Gamma(\ell + 2)}{4\sqrt{\ell + \mbox{$\frac{3}{2}$}}\,
\Gamma\left(\ell + \mbox{$\frac{3}{2}$}\right) - \kappa\,\Gamma(\ell + 1)}}
\quad ,
$$
which, in turn, implies for the (variational) effective energy eigenvalues of
the funnel potential
$$
E_{\nb{eff}}(\lambda_{\nb{min}}) = 2\,\lambda_{\nb{min}}
\left(3\sqrt{\ell + \mbox{$\frac{3}{2}$}} - \kappa\,\dfrac{\Gamma(\ell + 1)}
{\Gamma\left(\ell + \mbox{$\frac{3}{2}$}\right)}\right) \quad .
$$

The large-$\ell$ behaviour of the ultrarelativistic (variational) effective
energy eigenvalues $E_{\nb{eff}}(\lambda_{\nb{min}})$ resulting from this
expression is the same as for the pure linear potential:
$E_{\nb{eff}}^2(\lambda_{\nb{min}}) = 9\,a\,\ell$. This circumstance is an
unavoidable consequence of the fact that in the limit $\ell \to \infty$ all
contributions of the Coulomb part of the funnel potential to the above
effective energy eigenvalues vanish, which may be seen immediately by
recalling once more the above-mentioned asymptotic behaviour of the ratio of
gamma functions. Accordingly, for very large orbital angular momenta $\ell$
the positioning of the energy levels of the funnel potential is controlled by
its confinement part only.\footnote{\normalsize A similar observation has
already been made in Ref. \cite{lucha91regge} within a slightly different
context \cite{lucha91rel,lucha90com}.}

Table \ref{tab:funnel} compares the nonrelativistic and effectively
semi-relativistic approaches, as before, and, in addition, also the truly
semi-relativistic treatment for the funnel potential. The results for both
nonrelativistic and effectively semi-relativistic approach have been computed
by just the same numerical procedure \cite{falk85} as before whereas the
results for the semi-relativistic treatment have been obtained, also only on
numerical footing, by the so-called \ao method of orthogonal collocation"
\cite{orthcoll}. This method tries to approximate the action of the
square-root operator of the relativistically correct kinetic energy,
$\sqrt{\vec p\,{}^2 + m^2}$, on some suitably chosen (truncated) set of basis
states by a well-defined (finite) matrix representation. For our choice of
parameter values, in particular, for a mass $m$ of the bound-state
constituents as large as $m = 1.8\mbox{ GeV}$, we find that the effective
approach is worse than the nonrelativistic one: $E_{\nb{SR}} < E_{\nb{NR}} <
E_{\nb{eff}}$ for all states.

{\normalsize
\begin{table}[hbt]
\begin{center}
\caption{Energy eigenvalues (in GeV) of nonrelativistic, effectively
semi-relativistic, and truly semi-relativistic Hamiltonian for the funnel
potential $V(r) = - \kappa/r + a\,r $, with $\kappa = 0.456$ and $a = 0.211
\mbox{ GeV}^2$.}\label{tab:funnel}
\vspace{0.5cm}
\begin{tabular}{|c|c|c|c|}
\hline
&&&\\[-1ex]
\multicolumn{1}{|c|}{State}&
\multicolumn{1}{c|}{\begin{tabular}{c}
Nonrelativistic\\ Hamiltonian\end{tabular}}&
\multicolumn{1}{c|}{\begin{tabular}{c}
Effective\\ Hamiltonian\end{tabular}}&
\multicolumn{1}{c|}{\begin{tabular}{c}
Semi-relativistic\\ Hamiltonian\end{tabular}}\\
&&&\\[-1.5ex]
\hline
&&&\\[-1.5ex]
$\quad\mbox{1S}\quad$&$\quad$3.9655$\quad$&$\quad$4.0172$\quad$
 &$\quad$3.9250$\quad$\\
$\quad\mbox{1P}\quad$&$\quad$4.4014$\quad$&$\quad$4.4471$\quad$
 &$\quad$4.3707$\quad$\\
$\quad\mbox{2S}\quad$&$\quad$4.5873$\quad$&$\quad$4.6378$\quad$
 &$\quad$4.5217$\quad$\\[1.5ex]
\hline
\end{tabular}
\end{center}
\end{table}}

Table \ref{tab:massdep}, however, by diminishing step by step the numerical
value of the mass $m$ of the bound-state constituents while retaining at the
same time the numerical values of the potential parameters $\kappa$ and $a$,
demonstrates by the example of the obtained ground-state energy level that
this situation may change drastically when reducing the value of $m$. Below
$m \simeq 0.75\mbox{ GeV}$ we find (for the 1S state) $E_{\nb{SR}} <
E_{\nb{eff}} < E_{\nb{NR}}$. These inequalities indicate that for
comparatively small masses $m$ the effective treatment represents a better
approximation to the true, i.~e., semi-relativistic, energy eigenvalues than
the nonrelativistic approach.

{\normalsize
\begin{table}[hbt]
\begin{center}
\caption[]{Dependence of the ground-state (1S) energy eigenvalues (in GeV) on
the mass $m$ of the bound-state constituents within nonrelativistic,
effectively semi-relativistic, and truly semi-relativistic approach for the
funnel potential. The parameter values used here are the same as in Table
\ref{tab:funnel}.}\label{tab:massdep}
\vspace{0.5cm}
\begin{tabular}{|c|c|c|c|}
\hline
&&&\\[-1ex]
\multicolumn{1}{|c|}{$m$ [GeV]}&
\multicolumn{1}{c|}{\begin{tabular}{c}
Nonrelativistic\\ Hamiltonian\end{tabular}}&
\multicolumn{1}{c|}{\begin{tabular}{c}
Effective\\ Hamiltonian\end{tabular}}&
\multicolumn{1}{c|}{\begin{tabular}{c}
Semi-relativistic\\ Hamiltonian\end{tabular}}\\
&&&\\[-1.5ex]
\hline
&&&\\[-1.5ex]
$\quad\mbox{0.250}\quad$&$\quad$1.6677$\quad$&$\quad$1.5712$\quad$
&$\quad$1.4078$\quad$\\
$\quad\mbox{0.336}\quad$&$\quad$1.6996$\quad$&$\quad$1.6540$\quad$
&$\quad$1.5066$\quad$\\
$\quad\mbox{0.500}\quad$&$\quad$1.8539$\quad$&$\quad$1.8524$\quad$
&$\quad$1.7256$\quad$\\
$\quad\mbox{0.750}\quad$&$\quad$2.1904$\quad$&$\quad$2.2149$\quad$
&$\quad$2.1054$\quad$\\
$\quad\mbox{1.000}\quad$&$\quad$2.5807$\quad$&$\quad$2.6168$\quad$
&$\quad$2.5161$\quad$\\[1.5ex]
\hline
\end{tabular}
\end{center}
\end{table}}

Table \ref{tab:errors}, by confronting the relative deviations of the
nonrelativistic ($E_{\nb{NR}}$) and effectively semi-relativistic
($E_{\nb{eff}}$) energy eigenvalues from their proper semi-relativistic
($E_{\nb{SR}}$) counterparts (normalized to $E_{\nb{SR}}$), works out this
tendency still more clearly: for masses $m$ of the bound-state constituents
restricted (for the present set of parameter values) by $m < 0.75\mbox{ GeV}$
the relative errors of our effective treatment become smaller than those of
the nonrelativistic one. (A further, rather trivial lesson to be learned from
the relative differences of energy eigenvalues compared by Table
\ref{tab:errors} reads: the heavier the bound-state constituents are, the
more irrelevant the precise form of the employed kinematics becomes and,
consequently, the more the results of the different ways of description will
approach each other---the energy of the bound state is dominated by the sum
of the masses of the constituents.)

{\normalsize
\begin{table}[hbt]
\begin{center}
\caption[]{Relative differences of the energy eigenvalues of the (1S) ground
state within nonrelativistic, effectively semi-relativistic, and truly
semi-relativistic treatment of the funnel potential. The parameter values
used here are the same as in Table \ref{tab:funnel}.}\label{tab:errors}
\vspace{0.5cm}
\begin{tabular}{|c|r|r|}
\hline
&&\\[-1ex]
\multicolumn{1}{|c|}{$m$ [GeV]}&
\multicolumn{1}{c|}
{$\dfrac{E_{\mbox{\small NR}} - E_{\mbox{\small SR}}}
{E_{\mbox{\small SR}}}$ [\%]}&
\multicolumn{1}{c|}
{$\dfrac{E_{\mbox{\small eff}} - E_{\mbox{\small SR}}}
{E_{\mbox{\small SR}}}$ [\%]}\\
&&\\[-1.5ex]
\hline
&&\\[-1.5ex]
$\quad\mbox{0.250}\quad$&18.5$\quad\qquad$&11.6$\quad\qquad$\\
$\quad\mbox{0.336}\quad$&12.8$\quad\qquad$&9.8$\quad\qquad$\\
$\quad\mbox{0.500}\quad$&7.4$\quad\qquad$&7.3$\quad\qquad$\\
$\quad\mbox{0.750}\quad$&4.0$\quad\qquad$&5.2$\quad\qquad$\\
$\quad\mbox{1.000}\quad$&2.6$\quad\qquad$&4.0$\quad\qquad$\\
$\quad\mbox{1.800}\quad$&1.5$\quad\qquad$&2.6$\quad\qquad$\\[1.5ex]
\hline
\end{tabular}
\end{center}
\end{table}}

\section{Conclusions}\label{sec:effhamconcl}

The present work has been dedicated to the formulation of effectively
semi-relativistic Hamiltonians which are designed in such a way that---by
suitable interpretation of their (effective) parameters---they allow us to
approximate an entirely semi-relativistic formalism at a formally
nonrelativistic level. Application of the developed formalism to a few
representative static interaction potentials gave indications that below some
specific critical mass of the involved particles, where the obtained energy
eigenvalues are closer to the (exact) semi-relativistic ones than those of a
nonrelativistic description, the effective approach represents, at least in
relativistic regions, an improvement of the certainly rather crude
nonrelativistic approximation. Simultaneously, this observation might
contribute to the eventual explanation of the surprising success of (a
variety of) nonrelativistic potential models in describing hadrons as bound
states of quarks even for the case of relativistically moving constituents.

\end{document}